%% file: sdolanNuPhysProceedings_arxive.tex
\documentclass[12pt]{article}
\usepackage{graphicx}
\usepackage{amsmath}
\usepackage{amsfonts}
\usepackage{amssymb}
\usepackage{graphicx} 
\usepackage{float} 
\usepackage{setspace} 
\usepackage{wrapfig}
\usepackage{gensymb}
\usepackage{lineno,xcolor}

\newcommand\pubnumber{NuPhys2015-Dolan}
\newcommand\pubdate{\today}

\def\oxford{Department of Physics,\\
Denys Wilkinson Building,\\ University of Oxford,\\ Oxford, OX1 3RH,\\ United Kingdom}
\def\support{\footnote{s.dolan@physics.ox.ac.uk}}

\def\Title#1{\begin{center} {\Large #1 } \end{center}}
\def\Author#1{\begin{center}{ \sc #1} \end{center}}
\def\Address#1{\begin{center}{ \it #1} \end{center}}

\newcommand\pubblock{\rightline{\begin{tabular}{l} \pubnumber\\
         \pubdate  \end{tabular}}}
\newenvironment{Abstract}{\begin{quotation}  }{\end{quotation}}
\newenvironment{Presented}{\begin{quotation} \begin{center} 
             PRESENTED AT\end{center}\bigskip 
      \begin{center}\begin{large}}{\end{large}\end{center} \end{quotation}}

\input econfmacros.tex
\let\OLDthebibliography\thebibliography
\renewcommand\thebibliography[1]{
  \OLDthebibliography{#1}
  \setlength{\parskip}{0pt}
  \setlength{\itemsep}{0pt plus 0.3ex}
}

\begin{document}

\begin{titlepage}
\pubblock

\vfill
\Title{Probing Nuclear Effects at the T2K Near Detector Using Transverse Kinematic Imbalance}
\vfill
\Author{Stephen Dolan\support \\ On behalf of the T2K collaboration}
\Address{\oxford}
\vfill
\begin{Abstract}
In this work we utilise variables characterising kinematic imbalance in the plane transverse to an incoming neutrino, which have recently been shown to act as a direct probe of nuclear effects (such as final state interactions, Fermi motion and multi-nucleon processes) in $\mathcal{O}$(GeV) neutrino scattering. We present a methodology to measure the charged current differential cross-section with no final state pions and at least one final state proton ($CC0\pi+Np, N \geq 1$) in these variables at the near detector of the T2K experiment (ND280), using the upstream Fine Grained Detector (FGD1) as a hydrocarbon target. Overall these measurements will allow us to better understand the impact of nuclear effects on the observables in neutrino scattering, providing valuable constraints on the systematic uncertainties associated with neutrino oscillation and scattering measurements for both T2K and other experiments with similar energy neutrino beams.
\end{Abstract}
\vfill
\begin{Presented}

NuPhys2015, Prospects in Neutrino Physics

Barbican Centre, London, UK,  December 16--18, 2015

\end{Presented}
\vfill
\end{titlepage}
\def\thefootnote{\fnsymbol{footnote}}
\setcounter{footnote}{0}

\vspace{-3mm}

\section{Nuclear Effects and Single Transverse Variables}

In $\mathcal{O}$(GeV) neutrino-nucleon scattering the kinematics of the observed final state depends on both the interaction type and various nuclear effects. It is therefore essential to understand these effects in order to accurately reconstruct the incoming neutrino energy, as is required in neutrino oscillation analyses. Nuclear effects can broadly be split into initial state effects, including Fermi motion (FM), nuclear binding energy and multi-nucleon processes (MNP), and final state interactions (FSI). FM describes the momentum distribution of nucleons prior to any interaction whilst MNP describe correlations between nucleons and are thought to be dominated by 2p2h effects for the neutrino energies relevant to T2K~\cite{Nieves3, Martini1}. FSI are defined by secondary interactions inside the nuclear target, including pion absorption and emission as well as the ejection of additional nucleons.  

\begin{figure}[h]
\centering
\includegraphics[scale=0.38]{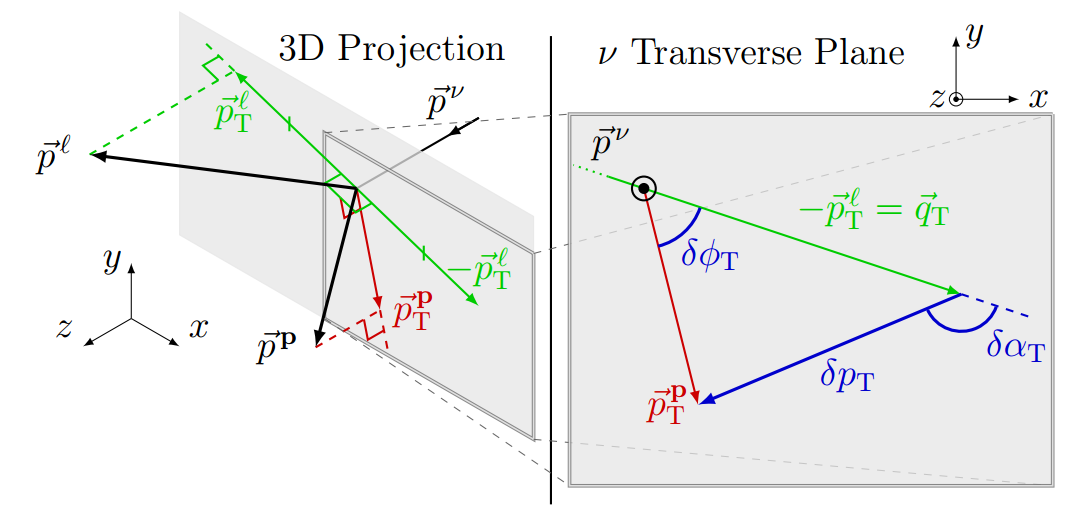}
\caption{ \small An illustration of a $CC0\pi+1p$ interaction with nuclear effects to provide a schematic definition of the three single transverse variables, $\delta \phi_\textrm{T}$, $\delta p_\textrm{T}$ and $\delta \alpha_\textrm{T}$. Here $\vec{p}^{\,\nu}$ is the incoming neutrino momentum whilst $\vec{p}^{\,\ell/\textrm{P}}$ is outgoing lepton/proton momentum. The T subscript denotes projection onto the plane transverse to the neutrino. Taken from~\cite{luke} }
\label{fig::svtdef}
\end{figure}

In this work, we propose to probe nuclear effects via the transverse kinematic imbalance between the muon and the highest momentum final-state proton in muon neutrino scattering with 1 muon, $N$ proton(s) ($N>0$) and no pions as a final state topology ($CC0\pi+Np$). Without nuclear effects in a $CC0\pi+1p$ topology (which therefore contains only $CCQE$ interactions), the momentum of the outgoing muon, projected into the plane transverse to an incoming neutrino, is exactly equal and opposite to that of the proton. However, once nuclear effects and/or other topologies are considered, the transverse momenta are no longer balanced. By reversing the outgoing muons' transverse momentum a ``transverse triangle'' is formed, clearly defining a complete set of single transverse variables (STV) to characterise the transverse imbalance and therefore nuclear effects (see figure ~\ref{fig::svtdef}). This is discussed further in~\cite{stv, luke}. 

It has been shown that the STV can act as excellent probes of FSI and FM, whilst exhibiting only a minimal dependence on neutrino energy for exclusive final states~\cite{stv, luke}.  Therefore the STV potentially allow a deconvolution of flux uncertainties and nuclear effects. By using ND280's excellent tracking resolution to measure and unfold these variables we hope to make state of the art measurements of nuclear effects.

\vspace{-3mm}

\section{ND280 Detector and Event Selection} 
\label{sec::sel}

ND280 is placed $2.5 \degree$ off-axis in a neutrino beam, provided by the J-PARC accelerator facility, with a peak (off-axis) energy of about $0.6$ GeV. The flux is well constrained by hadron production data from the NA61/SHINE experiment~\cite{T2KFlux}. ND280's  FGD1 is used as a hydrocarbon target for neutrino interactions and both FGD1 and time projection chamber (TPC) 1 and 2 (all inside a 0.2 T magnetic field) are used for tracking. The tracker electro-magnetic calorimeter (ECal) is used as a veto for $\pi^0$ like events~\cite{T2K}.
 
An event selection is made using events generated by the NEUT Monte-Carlo (MC) simulation~\cite{NEUT} fed through an ND280 detector simulation. The NEUT MC generated in this study uses a relativistic Fermi gas FM model, the Nieves 2p2h model~\cite{Nieves3} and a cascade model to describe FSI. Events with one muon, at least one proton and no other identified particles are selected. These events are split into four signal topologies, divided based on which detectors are used for reconstructing the selected event and how many protons are detected. This is necessary to account for the different detector acceptances in each topology. Additionally, two control samples are used to constrain the dominant background processes (resonant and deep inelastic scattering charged current interactions).

To ensure a reasonable, and relatively flat, detector acceptance in each STV bin, the signal is defined as $CC0\pi+Np$ with the following constraints on the muon/proton momentum ($p_{\mu / \textrm{p}}$) and angle ($\theta_{\mu / \textrm{p}}$): $p_\mu>250$ MeV/\textit{c}, $p_\textrm{p}>450$ MeV/\textit{c}, $p_\textrm{p}<1000$ MeV/\textit{c}, $cos(\theta_{\mu})>-0.6$ and $cos(\theta_\textrm{p})>0.4$. After the section the total efficiency and purity for all $CC0\pi+Np$ events is 12.6\% and 83.5\% respectively, leaving 3630 events expected across the 4 signal topologies in current T2K neutrino mode data (corresponding to $5.811\times 10^{20}$ protons on target). After applying the phase space restrictions the efficiency raises to 30.0\%.

\begin{figure}[h]
\centering
\includegraphics[scale=0.185]{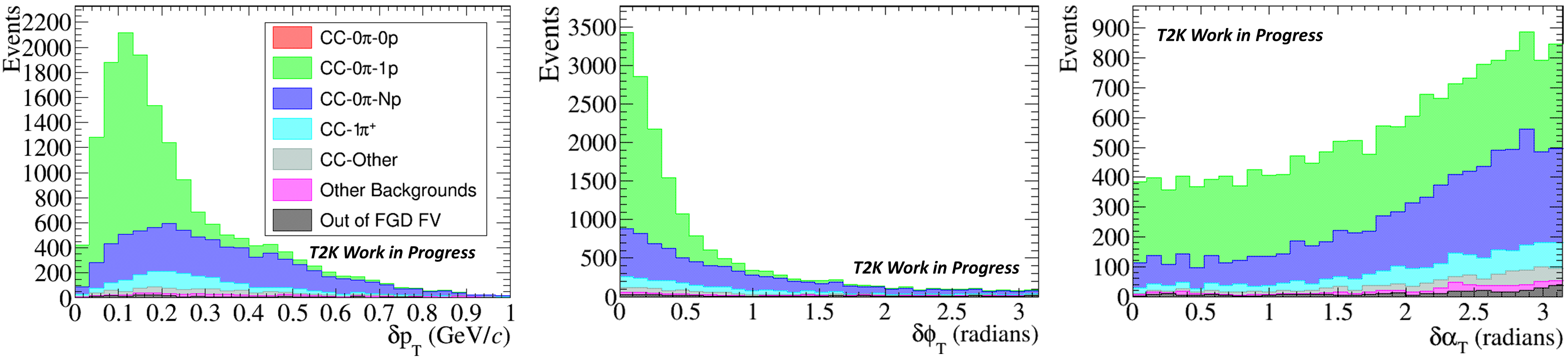}
\caption{ \small The STV distributions calculated using the truth kinematics of selected events, stacked by final state topology. Other backgrounds refers to neutral current, electron neutrino or anti neutrino scattering whilst FV stands for fiducial volume.}
\label{fig::selplots}
\end{figure}

In order to reconstruct the STV the initial neutrino direction is required. This is estimated as the direction of the vector from the mean neutrino parent decay point, in the decay tunnel of the T2K secondary beam line~\cite{T2KFlux}, to the interaction vertex position in the FGD. The muon and proton kinematics can then be used to form STV distributions of selected events, shown in figure~\ref{fig::selplots}. Note that the results shown are from a preliminary MC study, until this is finalised, in order to prevent bias, the study remains blind.   

\vspace{-2mm}

\vspace{-3mm}

\section{Cross Section Extraction} 

Since nuclear effects are poorly understood, it is important to make the cross section extraction method as model independent as possible. We therefore opt to use a regularised likelihood fit, which requires no preconceived notion of the distribution of selected events. The fit parameters are weights on reconstructed signal templates (where the signal is as defined in section~\ref{sec::sel}), each formed from a thin STV truth bin (such that the MC shape within the bin is relatively flat). The detailed method and treatment of systematic errors is the same as used in analysis 1 of~\cite{cc0pi} but for the addition of templates to cover the out of phase space contribution to the data and the following regularising penalty term, aimed to mitigate strong bin to bin anti-correlations:

\begin{equation}
\chi^2_{reg} = p_{reg} \sum_i (c_i-c_{i-1})^2
\end{equation}

Here $c_i$ is the weight of the template in reconstructed STV bins from the $i^{th}$ true STV bin, with respect to some prior (i.e. the fit parameters). $p_{reg}$ defines the strength of the regularisation and is chosen using the ``L-curve'' method described in~\cite{lcurve}.

\vspace{-3mm}

\section{Preliminary Fake Data Results} 
\label{sec::results}

\begin{figure}
\centering
\includegraphics[scale=0.35]{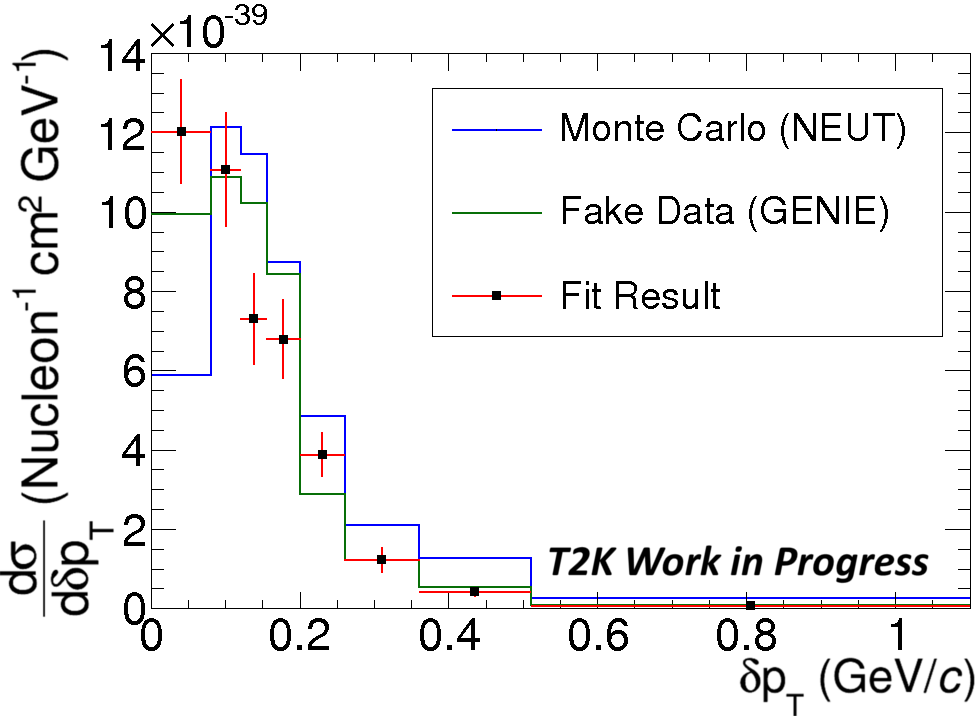}
\caption{ \small The fitted differential $CC0\pi+Np$ cross section (with phase space constraints listed in section~\ref{sec::sel}) in $\delta p_\textrm{T}$ is shown alongside the MC prediction and the fake data truth.}
\label{fig::xsec}
\end{figure}

To validate the fit procedure and show realistic expected sensitivity, we perform the analysis using NEUT as the MC and use the GENIE MC generator~\cite{GENIE} to produce a fake data set. The GENIE MC used here has no 2p2h and substantially different FSI modelling from NEUT. The results for $\delta p_\textrm{T}$ are shown in figure~\ref{fig::xsec}. Systematic parameters were included as nuisance parameters in the fit but the error bars show only the effect of statistical fluctuations expected with current T2K data. The total systematic error from all sources is expected to be around 20\%. Good agreement between the fake data and fit result (with 20\% uncorrelated systematic error $\chi^2/DoF=0.83$) and no obvious bias toward the nominal MC shows the success of the fit. A perfect fit cannot be expected since GENIE is not simply a systematic fluctuation of NEUT.

\vspace{-3mm}

\section{Conclusions and Further Work} 

Understanding nuclear effects in neutrino interactions is essential for making precision measurements of neutrino oscillations or exclusive cross sections. The STV provide a unique probe of these effects which, in some cases, is independent of neutrino energy. A methodology for measuring these distributions has been discussed and demonstrated for $\delta p_\textrm{T}$.

Future studies will finalise the fit procedure to allow T2K to extract a probe of nuclear effects with a minimal model dependence. Sensitivity to various effects will be evaluated by analysing the goodness of fit after turning off or changing elements of it (such as 2p2h, or the nuclear model used). An extension to this work will perform a model parameter fit to extract model dependent inputs for other cross section and oscillation analyses.

\end{document}

%% file: econfmacros.tex
\def\beq{\begin{equation}}
\def\eeq#1{\label{#1}\end{equation}}
\def\eeqn{\end{equation}}

\def\beqa{\begin{eqnarray}}
\def\eeqa#1{\label{#1}\end{eqnarray}}
\def\eeqan{\end{eqnarray}}

\let\bar=\overbar

\def\Dslash{\not{\hbox{\kern-4pt $D$}}}
\def\dslash{\not{\hbox{\kern-2pt $\del$}}}

\def\msb{{\bar{\ssstyle M \kern -1pt S}}}

